# Enhancement and tuning of the defect-induced electroluminescence of ZnO mesoporous layers in the visible range (preprint version)

Iännis Roland[1], Domitille Schanne[1], Alexandra Bogicevic[2] and Aloyse Degiron[1]

[1] Université de Paris, CNRS, Laboratoire Matériaux et Phénomènes Quantiques, F-75013 Paris, France
[2] Laboratoire de Physique et d'Etude des Matériaux, ESPCI-Paris, PSL Research University, CNRS UMR 8213, Sorbonne Université, 10 rue Vauquelin, F-75005 Paris, France

We show a way to pattern the visible electroluminescence of solution-processed mesoporous ZnO layers. Our approach consists in locally changing the nanoscale morphology of the coated ZnO layers by patterning the underlying surface with thin metallic patches. Above the metal, the ZnO film is organized in clusters that enhance its defect-induced electroluminescence. The resulting emission occurs over a large continuum of wavelengths in the visible and near-infrared range. This broad emission continuum is filtered by thin-film interferences that develop within the device, making it possible to fabricate LEDs with different colours by adjusting the thickness of their transparent electrode. When the metallic patterns used to change the morphology of the ZnO layer reach sub-micron dimensions, additional plasmonic effects arise, providing extra degrees of freedom to tune the colour and polarization of the emitted photons.

## 1. Introduction

Zinc Oxide (ZnO) is a choice material in optoelectronic research. An inherently n-doped semiconductor with a direct bandgap in the ultraviolet (UV) at 375 nm, ZnO has become the standard electron transport layer in quantum dot LEDs operating in the visible and infrared range [1–3]. Moreover, ZnO has also attractive luminescent properties by itself thanks to its large exciton binding energy that allows UV lasing at room temperature [4]. In particular, powders and polycrystalline layers of ZnO have been used to demonstrate random UV lasing [5–7]. ZnO under various forms (polycrystalline ZnO, nanowires, nanocrystals…) has also attracted considerable interest as an active material for UV LEDs, mostly in the form of n-doped regions of heterojunctions [8–13], but also in pure ZnO homojunctions thanks to progresses in p-type doping of this material [14].

Interestingly, ZnO also exhibit rich emission properties in the visible. The origin of this sub-bandgap emission is related to a variety of vacancy and interstitial defects that create a broad continuum spanning the whole visible range [15]. A potential application of this defect-induced emission is the development of white light emitting diodes, as suggested by studies in which ZnO nanocrystals (NCs) have been exploited as phosphors to convert UV or blue electroluminescence (EL) into a visible continuum [16,17]. In parallel, direct visible electroluminescence of vacancy and interstitial defects in ZnO has also been demonstrated by many teams [18–26]. As is the case for ZnO-based LEDs in the UV, most of these developments have been achieved by pairing an n-doped ZnO region (e.g. a polycrystalline or powder-based layer [19], a nanowire array [20–23], or a dense ensemble of ZnO colloidal NCs [24–26]) with a p-doped inorganic or organic layer, although it should be mentioned that ZnO homojunctions have also been demonstrated [18]. In these studies, the turn-on voltage is around 3-5 V and the EL spectra are typically measured at 8-10 V. It is noteworthy that external quantum efficiencies (EQE) are usually not reported although a recent study reports EQE of 4% at 6V [23]. In other words, the field of ZnO-based LEDs in the visible is not currently driven by performances but is a topic of academic interest. This fact is perhaps not surprising given that the origin of visible light emission in these devices comes from defects. Contrarily to band-to-band radiative recombination, which occurs within the whole medium, the visible light emission from ZnO is intrinsically limited by the density of defects. Moreover, such uncontrolled defects may also act as trap states that contribute to quench the radiative emission. In this context, there is an interest to develop strategies that maximize the defect density and their emissivity. A known avenue to enhance the defect-induced luminescence is to favor nano-composites (ZnO nanocrystals and nanowires…) over microstructured or bulk ZnO layers due to the higher surface-to-volume ratio [15] and most ZnO-based visible LEDs reported so far employ nanowires or nanocrystals [20–26].

In this study, we fabricate top-emitting LEDs operating with colloidal n-doped ZnO NCs in contact with p-doped polyvinylcarbazole (PVK), with indium tin oxide (ITO) as a transparent electrode. We show that the visible electroluminescence can be selectively enhanced by burying thin metallic patches under the surface of the ZnO NC assembly, increasing the granularity of the layer and thus the density of light-emitting defects. Several other levers, such as the ITO thickness and the geometry and metal of the underlying patches can be used to tune the colour emission and polarization in a controlled way.

## 2. Fabrication

Figure 1 shows the stack of the top-emitting LEDs investigated in this study. The substrate is a Si wafer covered with a 500 nm thick $SiO_2$ insulating layer. The fabrication begins with the deposition of 80 nm thick and 500 µm wide aluminum cathodes using a combination of photolithography on S1818 resist, metal evaporation and lift-off. Then, a solution of ZnO colloidal NCs from Avantama is coated onto the sample. The ZnO NCs have a diameter of 12 nm, their concentration is 25 g/l in isopropanol and they are spun at 3000 rpm to form a mesoporous layer with a thickness of about 50 nm. This layer is partially sintered on a hot plate bake at 200°C during 5 minutes to prevent erosion during further steps. The next step is the fabrication of 30 nm thick Au metallic patterns. In this study, various sizes are considered ranging from large patches covering hundreds of micron squares to sub-micron antennas. While the largest patterns can be defined by photolithography, we have relied most of the times on electron beam lithography (using a Raith Pioneer Two System and the CSAR 62 resist) so as to define in a single step all the features regardless of their size. The Au patterns are subsequently buried in ZnO by covering them with a monolayer (~12 nm) of ZnO NCs from Avantama. To obtain this thickness, the ZnO solution is diluted at 3 g/l in isopropanol and spin cast at 3000 rpm onto the sample. After a new bake at 100°C for 5 minutes, the hole injection layer in PVK is added to the stack. A thickness of about

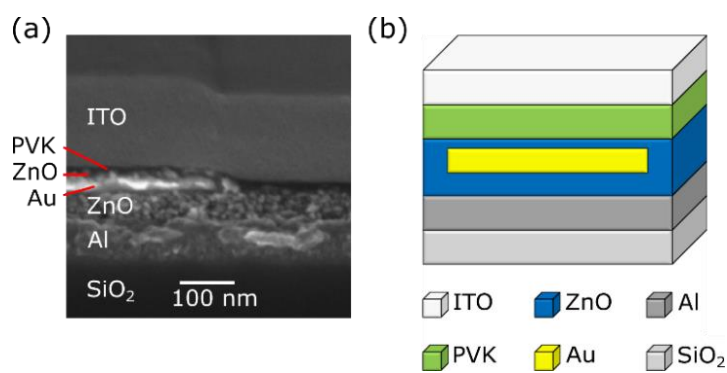

**Figure 1.** (a) SEM cross section view of the LED stack. (b) Schematic of a LED stack.

50 nm is obtained by spinning a solution of PVK from Sigma Aldrich at 8 g/L in chlorobenzene at 3000 rpm, which is then baked at 100°C during 5 minutes. The fabrication workflow ends with the deposition by radio-frequency sputtering of transparent ITO anodes through a shadow mask. As specified below, the thickness of the ITO anodes is set between 75 nm and 150 nm depending on the targeted emission colour. Because ITO is the last material deposited on the sample, we were not able to anneal it to improve its conductivity because it would have degraded the soft PVK layer underneath.

## 3. Observations

We start our study by examining a LED with a $50 \times 50$ µm² Au patch near its center. An optical image of the structure is given in Fig. 2(a). The horizontal stripe is the bottom aluminum cathode and the vertical stripe is the top ITO anode (with a thickness of 150 nm in this particular instance). The Au patch, which is slightly buried within the ZnO layer as described previously, is clearly visible near the middle of the junction. Figure 2(b) shows an image of the EL when a 9 V forward bias voltage is applied to the junction. We can see that EL mainly occurs above the gold patch while it is negligible everywhere else.

This behaviour is largely independent from the lateral dimensions of the Au features, as can be appreciated on Fig. 2(c) where Au patches of various sizes have been embedded in a single junction. In this case also, EL preferentially occurs above the Au patterns, with the same intensity and the same reddish hue as in Fig. 2(b). The EL spectrum of these samples, shown in Fig. 2(d) is centered on red wavelengths (660 nm). This EL peak is quite broad, covering most of the visible range and extending up to 800 nm (the UV emission from ZnO bandgap is not visible because it is outside of the sensibility range of our Flame spectrometer from Ocean Insight).

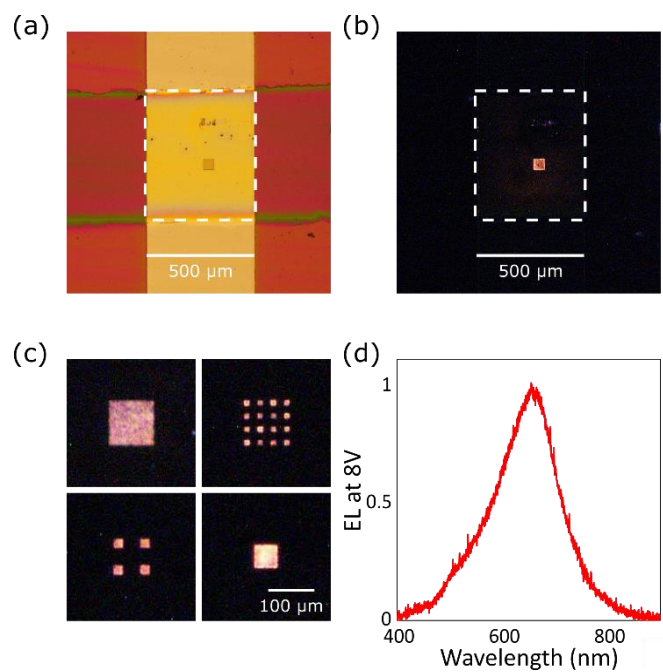

**Figure 2.** (a) Image of a junction seen through an optical microscope (magnification: 5X). (b) EL at 9 V captured by the same microscope. The dashed white square on both panels represents the limits of the LED, corresponding to the overlap between the horizontal Al cathode and the vertical ITO anode. (c) EL at 8 V from Au patches with various sizes. Although we display a composite image, all these patches are actually embedded within the same junction and the scale bar applies to all images. Upper left: array of 1024 Au squares with a side of 2 µm, period 3 µm (the individual squares are not visible at this magnification). Upper right: 16 squares with a side of 10 µm. Lower left: 4 squares with a side of 20 µm. Lower right: single square with a side of 50 µm. (d) EL spectrum at 8 V of the square Au patch with a 50 µm side observed on panel (c).

## 4. Origin of the enhancement

The behaviour evidenced on Fig. 2 only occurs when the Au patches are slightly buried under the ZnO surface. This result indicates that the source of EL is not Au itself: in particular, no EL enhancement has been observed when the Au patch was inserted at the interface between the n-doped ZnO and the p-doped PVK layers, which would have been in principle the best position for injecting electrons and holes within the Au region. In addition, photoluminescence experiments performed on half-finished samples with and without PVK reveal that the latter material does not contribute to the enhancement (results not shown here). These observations imply that the EL signal originates from the ZnO layer and that the brightness is considerably improved in the presence of the Au patches. However, we can rule out a contribution from the optical (plasmon) resonances of the Au patches because optical resonances are inherently geometry-dependent while we have seen that the EL colour and intensity do not depend on the size of the Au patches [Fig. 2(c)]. Likewise, it is also unlikely that Fermi pinning effects and band bending at the interface between ZnO and Au play a key role in the process: if such were the case, the preferential EL enhancement would be largely independent from the depth in which the Au patches are buried within ZnO. Besides, the results presented in Fig. 2 do not depend on the work function of the metal and the same results can also be reproduced with Ag, as will be shown in section 6.

To understand the origin of the EL enhancement, it is useful to inspect the surface of the second ZnO layer coated on the sample to bury the Au patterns. To this end, we have probed the samples with atomic force microscopy (AFM) before covering them with PVK and ITO. Figure 3(a) shows the topology of the ZnO surface on an area without Au patterns. A uniform granularity with feature sizes of approximately 10-15 nm is observed, which is consistent with the nominal diameter (12 nm) of the individual NCs forming this mesoporous layer. In contrast, Fig. 3(b) reveals that the surface morphology of the ZnO layer changes significantly above the Au patterns: while the image also features a clustering aspect, the size of the grains revealed by the AFM is about 50 nm in the plane of the sample and 30 nm in the direction normal to the surface. In other words, the organization of the ZnO NCs outside and above the Au patterns is quite different. This result is not surprising given that this second ZnO layer is very thin, implying that it is very sensitive to the morphology of the underlying material. In particular, the larger features seen on Fig. 3(b) are most likely related to the grains of the Au pattern underneath. In addition, a complementary inspection of the surface with a scanning electron microscope (SEM) suggests that this thin ZnO layer is not necessarily continuous above the Au patterns, as shown on Fig. 3(c).

These surface changes are most likely responsible for the EL enhancement: since ZnO can only emit visible light via defects, which are themselves strongly dependent on the morphology of the layer, it is fully expected from the literature [15] that different NC organizations result in different emission yields. In addition, and as stated earlier, the results observed on Fig. 2 critically depend on the position of the Au patterns within the stack of the LED—meaning in particular that no EL enhancement occurs if the ZnO layer covering the Au patterns increases to the point of forming a continuous mesoporous film.

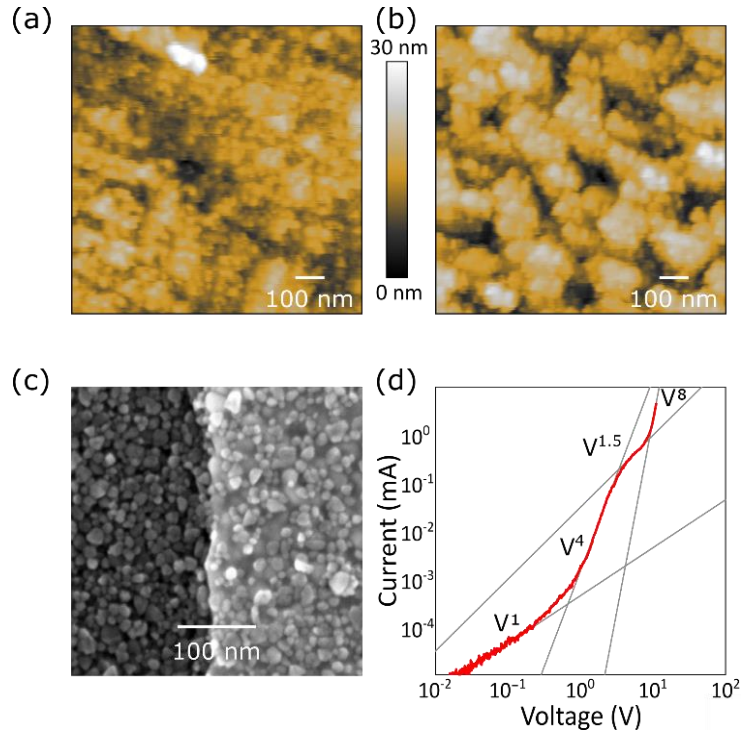

**Figure 3.** (a) Surface topography map of the second layer of ZnO NCs deposited onto the sample. This measurement was taken with a Core AFM from Nanosurf in a region of the sample without Au pattern. (b) Same measurement for the ZnO NCs above a Au patch. (c) SEM micrograph of the second ZnO layer coated onto the sample. The boundary between the two shades of grey corresponds to the left edge of a Au patch. (d) Current voltage characteristic of the device investigated on Figs. 2(c) and 2(d).

As a last evidence that the behaviour of the LEDs is dominated by the defect states in ZnO, we show the current-voltage (I-V) curve of a representative device in a log-log plot on Fig. 3(d) (only the forward characteristic is shown because our junctions fail when biased with negative voltages). The I-V curve cannot be fitted with the usual exponential function typical of a P-N junction, but rather with a succession of power laws with different exponents. This observation is indicative of a space-charge-limited current (SCLC), a regime already observed in several transport studies on ZnO [27–31]. In some instances, the value of the exponents used to fit the different parts of the curve, as well as the voltage ranges at which each power law occurs can be used to extract quantitative information on the nature and the density of the defect states [32]. However, such an analysis can only be performed for continuous layers where quantities such as the permittivity, length, and intrinsic semiconducting parameters can be defined, which is impossible with our samples due to the discontinuous nature of the ZnO film above the Au patterns.

## 5. Tuning the colour of the LEDs

So far, we have only considered LED stacks with 150 nm thick ITO anodes. The optical microscope images of Fig. 4(a) show that modifying the thickness of this transparent electrode by a few tens of nm has a strong impact on the colour of the emitted light, which can be tuned from the red to the green (the blue hue of the first image is an incorrect interpolation of the CCD camera used to capture the images). The corresponding EL spectra are plotted on Fig. 4(b).

We ascribe this colour tunability to a filtering effect, as schematized on Fig. 4(c): because ITO has a high index of refraction, thin film interferences can develop within the stack even though the total thickness of the LED is small compared to the emission wavelengths. This hypothesis is confirmed by the reflectivity spectra of the samples plotted on Fig. 4(d): each curve exhibits an interference dip that shifts to the blue as the ITO thickness decreases from 150 nm to 75 nm, modulating the broadband defect-induced emission of the ZnO layer. The exact wavelengths of the reflectance dips do not strictly follow the EL peaks of Fig. 4(b). In particular, for the two thinnest samples, the reflectance dips are more shifted to the blue than the EL. This observation indicates that little light is emitted at these wavelengths even in the absence of thin film interferences. It also explains why almost no light is emitted by the thinnest sample: the latter filters most wavelengths of the visible range, except in the blue where the ZnO-induced emission is negligible.

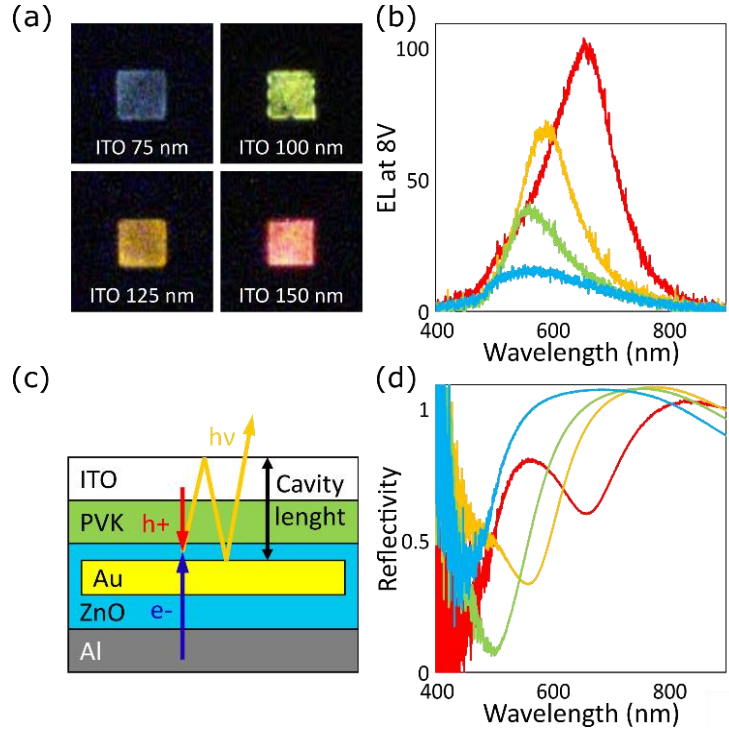

**Figure 4.** (a) Preferential EL above single Au squares with a side of 50 µm, for samples with 75 nm, 100 nm, 125 nm and 150 nm thick ITO anodes, respectively. Images taken with a 5X microscope objective. (b) EL spectra at 8 V from the same series. (c) Schematic of the thin film interferences creating the different colours (d) Reflectivity of the four samples measured above the Au patches.

## 6. From micro-patches to nano-antennas

So far, we have worked with non-resonant metallic micron-sized patches. In this last section, we briefly examine the case of resonant plasmonic inclusions. Because Au is not a good plasmonic material for wavelengths below 600 nm, our starting point is the same geometry as before, except that Au is replaced by Ag. Fig. 5(a) displays the EL produced by LEDs with $50 \times 50$ µm$^2$ and $10 \times 10$ µm$^2$ patches. We observe the same effects as those previously reported for Au, namely, a preferential emission above the metallic features, with a colour and intensity that do not depend on the size of the metallic patches.

We have then repeated the experiments with Ag inclusions of sub-micron dimensions that sustain plasmonic resonances in the visible range. The dimensions of the Ag particles can be appreciated on the SEM image of Fig. 5(b): they have an elongated shape with a length of 100 nm, a width of 40 nm, a thickness of 30 nm and they are arranged in a regular array with a periodicity of 200 nm. The reflectivity measurements plotted on Fig. 5(b) show that the sample sustain plasmon resonances at two different wavelengths in the visible range: one resonance polarized along the short axis at 520 nm, and another one polarized along the long axis at 700 nm.

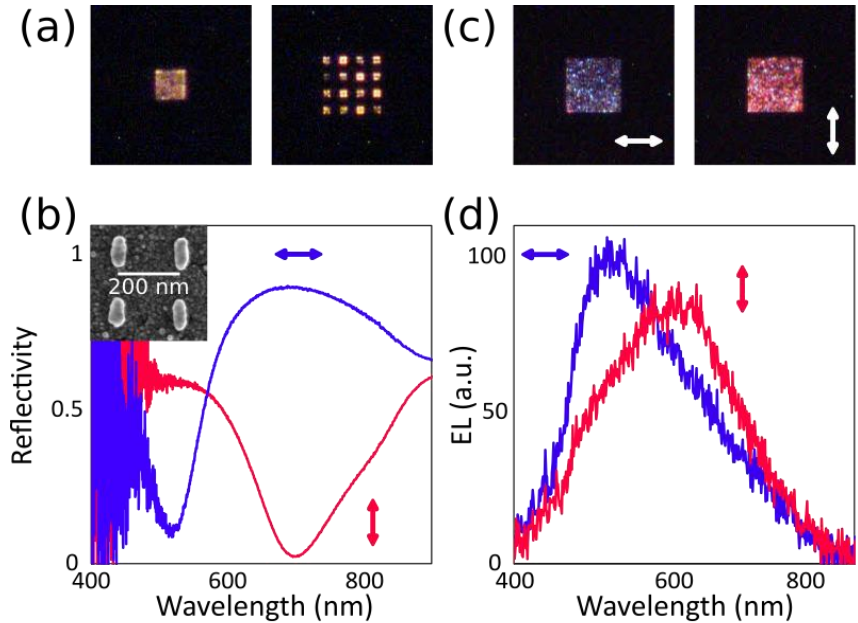

**Figure 5.** (a) EL of samples with micron-sized Ag patches. Left: single Ag square with a side of 50 µm. Right: matrix of 16 Ag squares with a side of 10 µm. (b) Reflectance spectra of a sample with 40 × 100 nm² Ag nano-antennas for two different polarizations (blue curve: polarization parallel to the short axis, red curve: polarization parallel to the long axis). The total size of the array is 100 × 100 µm². Inset: SEM image of the sample before coating the antennas with the second layer of ZnO NCs. (c) EL produced by the sample (left image: polarization parallel to the short axis, right image: polarization parallel to the long axis). (d) EL spectra for the same array and the same two polarizations.

The EL of such a device, visualized on Fig. 5(c) through a linear polarizer successively aligned along the short and long axes of the Ag particles, is markedly different from the emission above the micro-patches of Fig. 5(a). The emission is polarized, with a bluish hue for the horizontal polarization along the short axis of the particles and pink-red tones for the polarization perpendicular to it. As was the case for Fig. 4(a), these colours are somewhat misleading because the colour rendition of our camera favors the blue. The EL spectra, displayed on Fig. 5(d), show that the emission is actually green for the horizontal polarization and red for the vertical polarization. This polarization-dependent EL can be attributed to the coupling of the emitted light to the plasmon resonances of the Ag inclusions, which thus play the role of optical nano-antennas. As was the case with the thin-film interference effects of Fig. 4, the EL peaks do not strictly coincide with the maxima of the plasmon resonances evidenced on the reflectivity spectra of Fig. 5(b) because the EL spectra result from a convolution of the resonances and the bandwidth of the ZnO defect-induced emission.

## 7. Conclusion

In conclusion, we have created electroluminescent patterns within LEDs that emit visible light via the vacancy and interstitial defects of ZnO NC assemblies. These patterns were obtained by burying thin metallic shapes within the ZnO layer, locally enhancing the luminescence by modifying the organization of the ZnO NCs coated above them. Additional control of the light emission (colour and polarization) has been demonstrated using thin film interferences and by making the metallic features resonant. These results contribute to the emerging toolbox aiming at rendering ZnO-based LEDs competitive for visible and white light applications.